\documentclass[aps,prb,twocolumn]{revtex4}
\usepackage[a4paper,left=2cm,top=2cm,right=2cm,nohead,nofoot]{geometry}
\usepackage{epsfig}
\usepackage{graphicx}
\begin{document}

\title{Coexistence of tetrahedral and octahedral-like sites in amorphous phase change materials}
\author{S. Caravati$^1$,  
M. Bernasconi$^1$, T. D. K\"uhne$^2$, 
M. Krack$^2$, and M. Parrinello$^2$}

\affiliation{$^1$Dipartimento di Scienza dei Materiali,
 Universit\`a di  Milano-Bicocca, Via R. Cozzi 53, I-20125, Milano, Italy}

\affiliation{$^2$Computational Science, Department of Chemistry and
Applied Biosciences, ETH Zurich, USI Campus,
Via Giuseppe Buffi 13, 6900 Lugano, Switzerland} 

\begin{abstract}
Chalcogenide alloys are materials of interest for optical recording and  non-volatile memories.
We perform ab-initio molecular dynamics simulations aiming at shading light onto  the structure of amorphous
Ge$_2$Sb$_2$Te$_5$ (GST), the prototypical material in this  class. 
First principles simulations show that  amorphous GST obtained by quenching from the liquid phase displays two
types of short range order. One third of Ge atoms are in a tetrahedral environment while the remaining Ge, Sb and
Te atoms display a defective octahedral environment, reminiscent of cubic crystalline GST.
\end{abstract}
\pacs{???}

\maketitle

Phase change materials based on chalcogenide alloys are presently used in optical storage devices (DVD) and
 are promising materials for non-volatile electronic memories\cite{lacaita}. 
Both applications rely on the reversible
 and fast transition between the amorphous and crystalline phases which have 
different optical and electronic properties. Among the chalcogenide glasses,
 Ge$_2$Sb$_2$Te$_5$ (GST) is the material of choice for non-volatile memory applications due to its superior
performances in terms of speed of transformation and stability of the amorphous phase \cite{lacaita}. 

 In spite of its great technological importance, the microscopic structures of amorphous GST (a-GST)
 and the detailed mechanism of the phase transformation are largely unknown \cite{greer}. 
In the past
the structure of 
 a-GST was implicitly assumed to be a disordered version of the metastable cubic (rocksalt) 
crystalline geometry. Very recently 
based on EXAFS/XANES
 measurements,
Kolobov {\sl et al.} \cite{kolobov1} proposed
 that in a-GST Ge is tetrahedrally coordinated as opposed to its octahedral coordination
 in the crystalline phases (hexagonal and metastable cubic). 
Based on ab-initio calculations We{\l}nic {\sl et al} \cite{abinitio} proposed a spinel-like geometry for the
local structure of a-GST.
However, the model of  a-GST proposed by Kolobov {\sl et al} is in contrast with
other interpretation of EXAFS data \cite{lukovsky} and with
more recent Reverse Monte-Carlo (RMC) models fitted to XRD data \cite{RMC}.
On the other hand, models produced by RMC are subject to large uncertainties when fitted to the total
scattering function only, which for a-GST is the weighted sum of six partial pair correlation functions.
Further investigations are required to obtain a more compelling characterization of  a-GST
 which would facilitate the search for better performing materials. 

In this respect, first principles atomistic simulations can provide precious insight.
On the basis of successful previous studies on the crystalline phases
 \cite{abinitio,abinitio1,wuttig2} it is to be expected that Density Functional Theory  works  well also 
for the amorphous GST investigated  here.

Ab-initio Molecular Dynamics simulations have been performed using the scheme of K\"uhne {\sl et
al}\cite{asap}. In the spirit of the Car-Parrinello (CP) approach the wavefunction is not self-consistently
optimized during the dynamics. 
However, in contrast to CP, large integration time steps can be used in the simulation. This scheme leads to a
slightly dissipative dynamics of the type $-\gamma_D \dot{\bf R}_I$ where ${\bf R}_I$ are the ionic
coordinates. In Ref.~\onlinecite{asap} it is shown how to compensate for this dissipation and obtain a correct
canonical sampling. This scheme has been implemented in the CP2K suite of programs \cite{quickstep1,quickstep2}.
We use the PBE exchange correlation functional\cite{PBE} and Goedecker-type pseudopotentials\cite{GTH}. The
Kohn-Sham orbitals are expanded in a TZVP Gaussian-type basis set and the charge density is expanded in a
planewave basis set with a cut-off of 100~Ry.

\begin{figure}[ht!]
 \begin{center}
  \epsfig{file=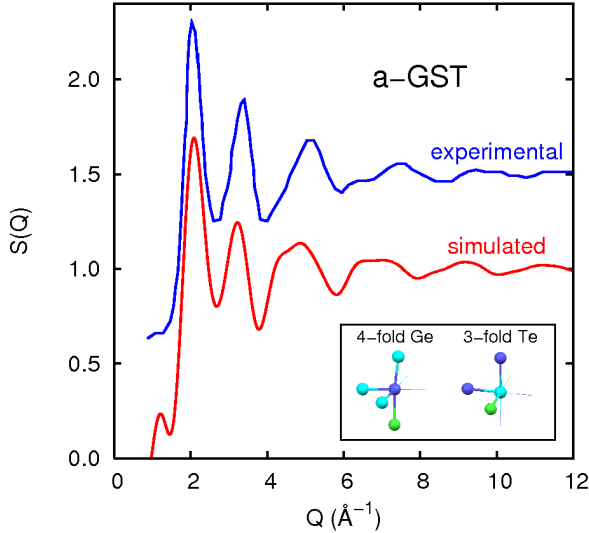,width=0.98\columnwidth}
 \end{center}
 \caption{(Color online) Calculated and experimental \cite{RMC} x-ray scattering factor $S(Q)$ of a-GST. The theoretical
$S(Q)$ has been computed from the partial structure factors $S_{ij}(Q)$ weighted by the $Q-$dependent x-ray
atomic form factors\cite{atformfact} ($S(Q)-1=F^X(Q)$ of Eq.~56 in Ref.~\onlinecite{corrfunct}). $S_{ij}(Q)$
are obtained in turn by Fourier transforming the partial pair correlation functions (3 ps at 300 K).
Inset: a sketch of the  geometry of defective octahedral sites
of 4-coordinated Ge and 3-coordinated Te.
}
 \label{xray-S_q}
\end{figure}

\begin{figure}[ht!]
 \begin{center}
 \end{center}
  \epsfig{file=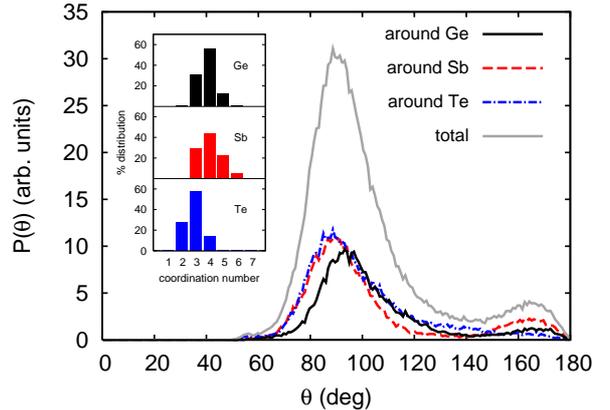,width=0.98\columnwidth}
 \caption{(Color online) Angle distribution function (total and resolved for different  central
atoms).  Inset: distribution of  coordination numbers of  different species obtained by integration of the
partial pair correlation functions (Fig.~S4 in EPAPS (Ref.~\onlinecite{ADD})).
}
 \label{angle}
\end{figure}

\begin{figure}[t]
 \begin{center}
 \end{center}
  \epsfig{file=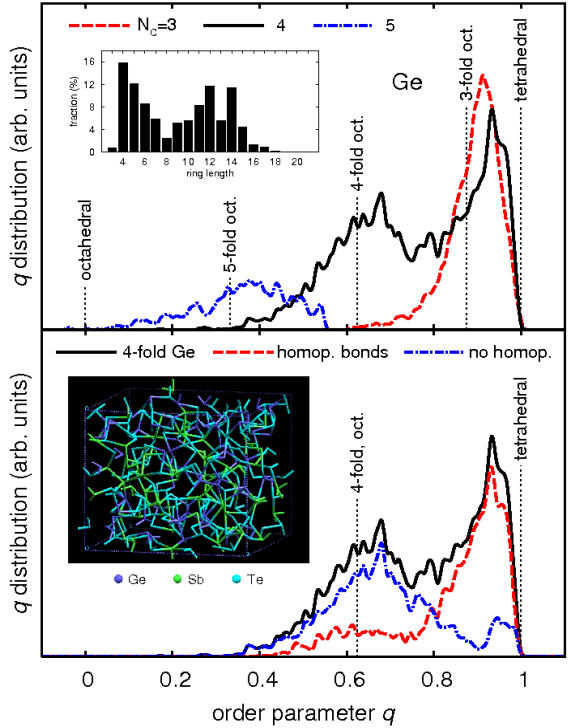,width=0.98\columnwidth}
 \caption{(Color online) Distribution of the local order parameter $q$ for Ge (see text).
$q$=1 for the ideal tetrahedral geometry, $q$=0 for the 6-coordinated octahedral site, and $q$=5/8 for a 4-coordinated
defective octahedral sites 
Top panel: $q$-distribution resolved for Ge with different
coordination number. Bottom panel: $q$-distribution for 4-coordinated Ge further resolved for Ge with at least
one homopolar bond (with Ge or Sb) or  bonding with Te only (no homopolar bonds). 
Top panel inset: ring distribution function of a-GST computed as in Ref.~\onlinecite{franz}.
Bottom panel inset: snapshot of the 270-atoms model
of a-GST.
}
 \label{qparam}
\end{figure}

The initial  configuration is the metastable cubic GST where  Te occupies
one sublattice of the rocksalt crystal and  Ge, Sb and vacancies are randomly placed in the other
sublattice in an orthorhombic supercell of size $21.97 \times 21.97 \times 18.63\ \rm\AA^3$ (270 atoms)
at the  density of 0.030 atoms /$\rm\AA^3$ close to the experimental
value for  a-GST\cite{density}.
 The system has then been heated and equilibrated for 6~ps  
at 2300~K and %, well above the experimental glass-transition temperature.
then quenched in 16 ps and
further equilibrated for 18 ps at %the temperature of
 990 K. %for which experimental diffraction data to compare with are available \cite{RMC}.
The parameter $\gamma_D=4\cdot10^{-4}$~fs$^{-1}$ has been determined as in the Ref.~\onlinecite{asap}.
The static properties of liquid GST turned out to be in good agreement with experiments \cite{RMC,neutronGST} as detailed in the
supplementary materials (Figs.~S1--S3 in EPAPS (Ref.~\onlinecite{ADD})).
In order to generate a model of a-GST the liquid has been brought to 300~K in 18~ps.
The calculated x-ray scattering function $S(Q)$
of a-GST reported  in
Fig.~\ref{xray-S_q} is in good agreement with XRD data \cite{RMC},
especially considering that the latter are on the as-deposited 
amorphous film which might be structurally slightly different 
from a-GST quenched for the melt as  suggested by their different  optical reflectivity
\cite{aprime}.
Incidentally, we recall that a-GST of relevance for applications is the 
phase quenched from the melt for which, to our knowledge,
 no XRD data are available. 
Average coordination numbers for the different species of a-GST (Table I) are computed from partial pair
correlation functions (Fig.~S4 in EPAPS (Ref.~\onlinecite{ADD})).
The distribution of the coordination number for the different species are reported in the inset of Fig.~\ref{angle}.
Ge and Sb atoms are mostly 4-coordinated and form bonds preferentially  with  Te atoms.
However, we observe a large fraction of homopolar Ge-Ge,  Sb-Sb and Ge-Sb bonds, namely
38 $\%$ of Ge are bonded with at least another Ge or Sb, a percentage which raises to 52 $\%$ for Ge
atoms 4-coordinated. The fraction of Sb bonded to another Sb or Ge is instead 59 $\%$.
Homopolar Ge-Ge bonds have been recently detected by x-ray fluorescence in cubic GST
grown epitaxially on crystalline GaSb \cite{fluo}.
The concentration of Te-Te bonds is somehow lower,  $ 27 \%$ of Te atoms are involved in
homopolar Te-Te bonds arranged into dimers and trimers (see Fig.~S5 in EPAPS (Ref.~\onlinecite{ADD})).
The large concentration of homopolar bonds is not reproduced by the RMC model
~\cite{RMC}.
 \begin{table}[ht!]
\begin{tabular}{|c|c|c|c|c|}
\multicolumn{5}{c}{}\\
\hline \hline
\multicolumn{5}{|c|}{\bf average coordination number} \\
\hline
 & with Ge & with Sb & with Te & total \\
\hline
Ge & 0.275 & 0.270 & 3.277 & 3.823 \\
Sb & 0.270 & 0.588 & 3.166 & 4.025 \\
Te & 1.311 & 1.267 & 0.288 & 2.866 \\
\hline \hline
\end{tabular}
 \caption{Average coordination number for different pairs of atoms computed from 
the partial pair correlation functions (Fig. S4 in EPAPS (Ref. \onlinecite{ADD})).}
 \end{table}
Insight on the local geometry is further gained from  the angle distribution function 
in Fig.~\ref{angle}. The broad peak at $\sim$90$^\circ$ and the weaker structure around $\sim$170$^\circ$ are reminiscent of the
distorted octahedral-like geometry of the metastable cubic crystal.
A snapshot of the a-GST model is shown in Fig.~\ref{qparam}. Angles at $\sim$90$^\circ$ and $\sim$180$^\circ$ clearly dominate the 
bonding network.
For Te, only angles at $\sim$90$^\circ$ are found (see inset of Fig.~\ref{xray-S_q}). 
 However, the main coordination of Ge(Sb) and
Te of four and three, respectively, is lower than  the ideal octahedral value of six.  
In our model the presence of neighboring vacancies is responsible for the lower coordination while the
bonding angles remain close to 
 $\sim$90$^\circ$ and $\sim$180$^\circ$   as in the metastable cubic phase.
 Te and Sb atoms adopt this configuration, while only a fraction of Ge atoms
are in a defective octahedral site. A large fraction of  4-coordinated Ge atoms are 
 in a tetrahedral environment as inferred from  EXAFS/XANES measurements \cite{kolobov1}.
A signature of the tetrahedral geometry is already visible as a shoulder at $\sim$110$^\circ$ in the
angle distribution function for Ge atoms in Fig.~\ref{angle}. A better indicator of the tetrahedral
geometry is given  by the local order parameter\cite{qdef} $q= 1-\frac{3}{8}\sum_{i > k} (\frac{1}{3} +
cos \theta_{ijk})^2$   where the sum runs over the couples of atoms bonded to a central atom $j$.
The distribution of the local order parameter $q$ for Ge atoms
is reported in Fig.~\ref{qparam} for  different  coordination numbers. The 4-coordinated
Ge  distribution is clearly bimodal with  peaks corresponding to  defective octahedra and 
tetrahedra. In contrast, the $q$-distribution for Te and Sb does not show any signature of
the tetrahedral geometry (Fig.~S6 in EPAPS (Ref.~\onlinecite{ADD})). 
The $q$-distribution for 4-coordinated Ge is further
analyzed in terms   of  atoms bonded to  Te only or to at least one Ge or Sb.
The presence of  bonds with Ge or Sb clearly favors the tetrahedral geometry.
Only few tetrahedral Ge are bonded to Te only. On the other hand, all Ge with more than one
homopolar bond (with Ge or Sb)  are in the tetrahedral geometry.
By integrating the tetrahedral peak of the $q$-distribution in the range 0.8-1.0 we estimate that
33 $\%$  
of  Ge  atoms are in a tetrahedral  environment.
The average bond length of Ge is slightly shorter  in tetrahedral sites (2.73 $\rm\AA$) than in
defective octahedral sites (2.82 $\rm\AA$, see Fig.~S6 in EPAPS (Ref.~\onlinecite{ADD})).
The presence of tetrahedral sites  and the absence of 6-coordinated octahedral sites 
(or distorted octahedra sites with 3+3 coordination as in crystalline $\alpha$-GeTe \cite{alfagete})  
is consistent with the
model inferred from EXAFS/XANES data in Ref.~\onlinecite{kolobov1}.  
The coexistence of tetrahedral and defective octahedral sites
found here might reconcile the different interpretation of EXAFS data \cite{kolobov1,lukovsky} based on 
the assumption of a unique environment 
for all Ge atoms.
The inclusion of both  configurations 
in a revised fitting might provide  a better agreement with experiments.
Turning now to the medium range order of a-GST, we report
the ring distribution in Fig.~\ref{qparam}. 
 Both even and odd rings are
present, as opposed to  the   RMC results ~\cite{RMC}.
The distribution has a pronounced maximum at four-membered rings typical of the rocksalt structure.
 The presence of very large rings (with very large aspect ratio)  reveals a somehow more open
and less connected structure than that of the  cubic crystalline phase.
%\clearpage

In summary, based on first principles simulations we have provided novel insight into the structure 
of amorphous Ge$_2$Sb$_2$Te$_5$ quenched from the melt. Most of Ge and Sb atoms in a-GST are 4-coordinated while Te is
mostly 3-coordinated in defective octahedral-like sites which recall the local environment of 
 cubic crystalline GST. 
 However, as many as 33 $\%$ of Ge atoms are
in a tetrahedral geometry, % peculiar of the amorphous phase
absent in the crystalline phase, and favored by the presence of homopolar (Ge-Ge and Ge-Sb) 
bonds. The coexistence of the two types of local environment is the key to understand the two
apparently contradictory and peculiar features of GST exploited in the devices, namely the strong optical (and perhaps electronic)
 contrast between the amorphous and crystalline structures 
and the high speed of the phase change.
This conclusion is also supported by the recent theoretical work  in Ref.~\onlinecite{abinitio1}) 
on models of amorphous GeSbTe alloys showing the presence of a fraction of Ge in tetrahedral
sites is indeed sufficient to produce a strong optical contrast between the amorphous and 
crystallie phases.

During the preparation of the manuscript we have received a preprint by Akola and Jones where simulations
similar to ours were reported. When applicable their results are similar to ours.  Computational resources have been
provided by CSCS and by CINECA through the CNISM-CNR program "Iniziativa Calcolo Parallelo 2007".  S.C.
acknowledges financial support given by Fondo Sociale Europeo, Ministero del Lavoro e della
Previdenza Sociale and Regione Lombardia. Discussion with R. Bez, A.
Modelli, A. Pirovano and E. Varesi are gratefully acknowledged.

\cleardoublepage

\renewcommand{\thefigure}{S1}
%\narrowtext
 \begin{figure}[h!]
%\begin{center}
% \epsfig{file=./figS1.eps,width=0.98\columnwidth}
%\end{center}
%\vskip -0.65cm
 \centerline{\includegraphics[width=0.98\columnwidth]{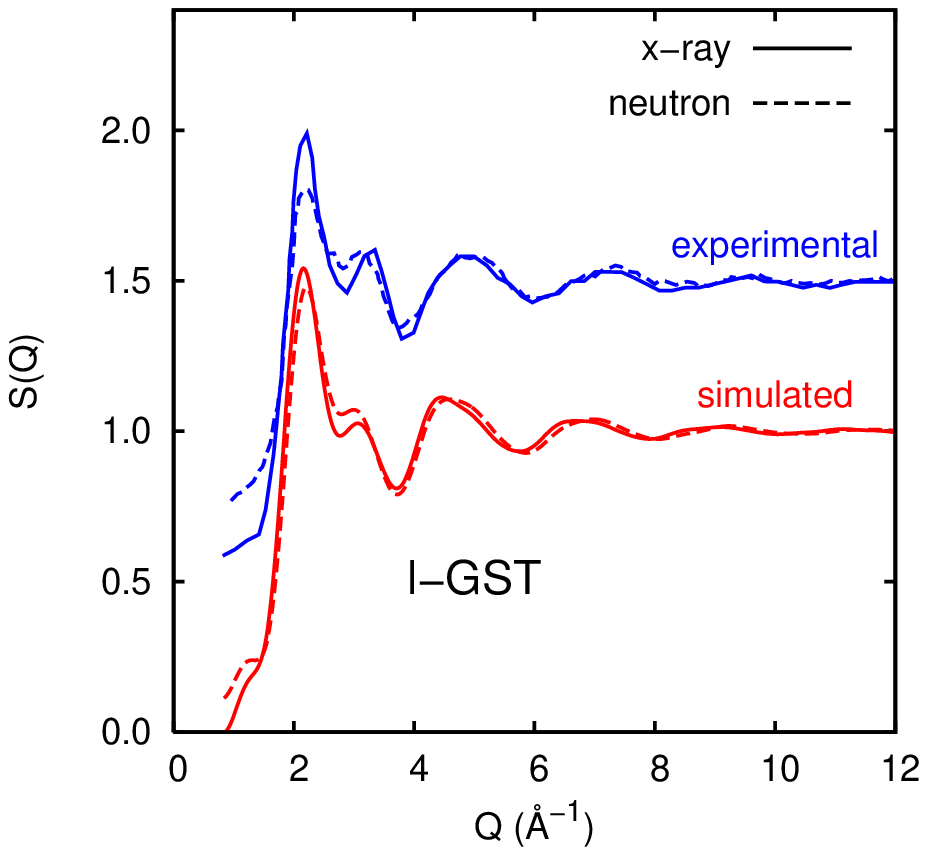}}
 \vskip -0.4cm
 \caption{Calculated and experimental scattering functions $S(Q)$ of liquid GST from x-ray (953~K, Ref.~6) and
neutron diffraction (1073~K, Ref.~15). $S(Q)$ has been calculated according to the definition by Keen (Ref. 23).
 Namely, $S(Q)$ is the sum of partial structure factors $S_{ij}(Q)$ weighted by $Q-$dependent atomic form factors
($S(Q)-1=F^X(Q)$ of Eq.~56 in Ref.~23) or by neutron scattering lengths (Eq.~19 in Ref.~23). $S_{ij}(Q)$ are
determined in turn by Fourier transforming the partial pair correlation functions $g_{ij}(r)$ (see Fig.~S2,
 990~K). \label{liquidSq}
}
\end{figure}
\renewcommand{\thefigure}{S2}
 \begin{figure}[h!]
%\begin{center}
% \epsfig{file=./figS2.eps,width=0.98\columnwidth}
%\end{center}
%\vskip -0.75cm
 \centerline{\includegraphics[width=0.98\columnwidth]{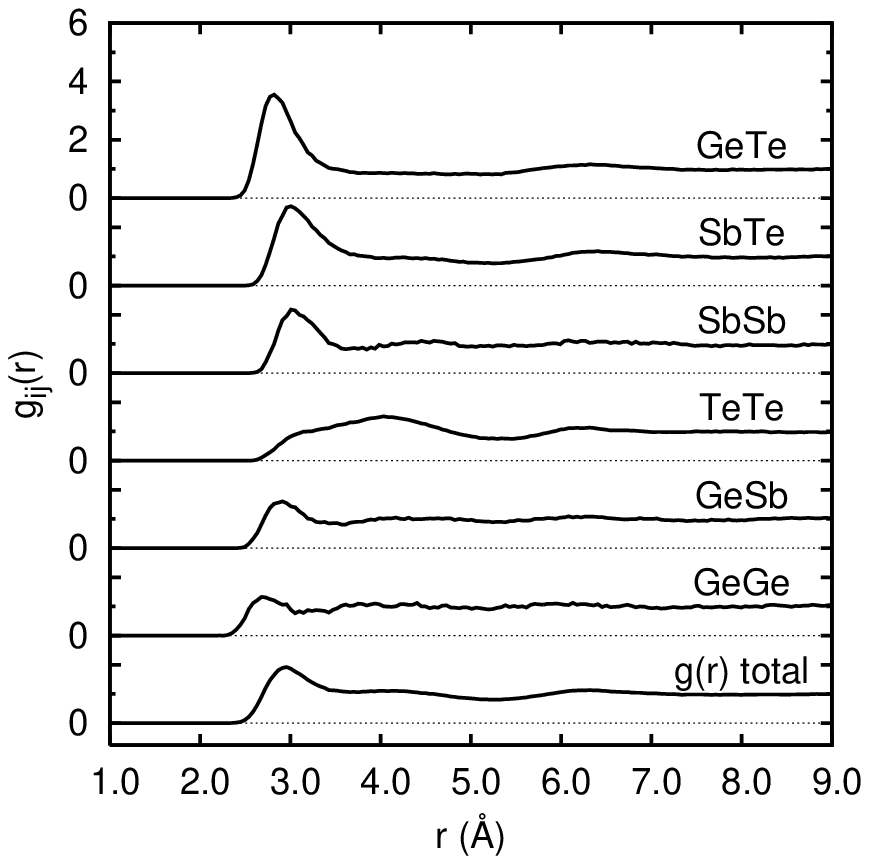}}
 \vskip -0.4cm
 \caption{Total and partial pair correlation functions of liquid GST (18 ps at 990 K). The first
peak in Sb-Sb  and Ge-Ge partial pair correlation functions are due to homopolar bonds. We have checked
that the system was liquid from the linear asymptotic growth of the mean square displacement. From the
Einstein relation we found $D=4.88\cdot10^{-5}$ cm$^2$/s. This is close to the estimate of $D=4.55 \cdot
10^{-5}$ cm$^2$/s obtained from shorter standard Born-Oppenheimer calculations performed to check the validity
of the $\gamma_D$ choice. To our knowledge no experimental data on the self-diffusion coefficient of liquid
GST are yet available.
}
 \label{lgofr}
\end{figure}
\renewcommand{\thefigure}{S3}
 \begin{figure}[h!]
%\begin{center}
% \epsfig{file=./figS3.eps,width=0.98\columnwidth}
%\end{center}
%\vskip -0.75cm
 \centerline{\includegraphics[width=0.98\columnwidth]{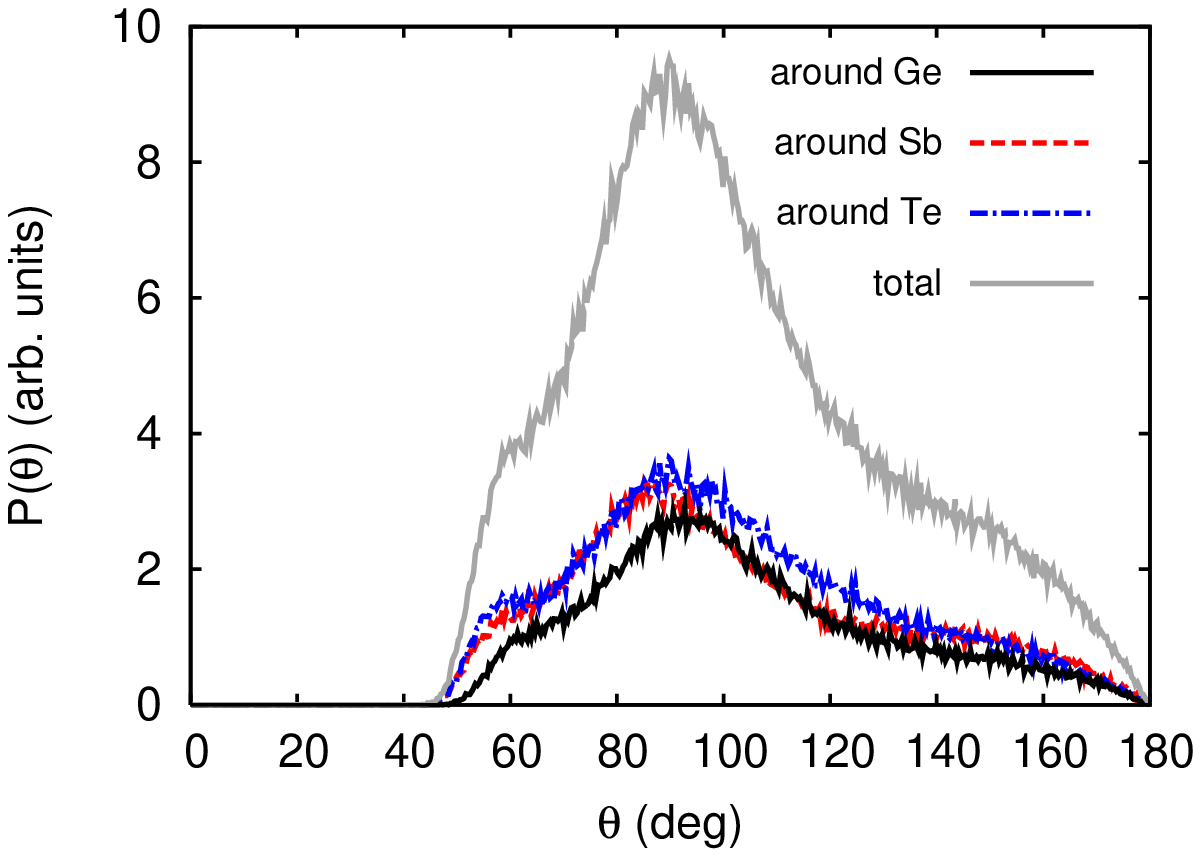}}
 \vskip -0.4cm
 \caption{Angle distribution functions of liquid GST at 990 K (total and resolved for
different central atoms). Angles at $60^\circ$ (absent in crystalline and amorphous GST)
correspond to triangular rings. \label{langle}
}
\end{figure}
\renewcommand{\thefigure}{S4}
 \begin{figure}[h!]
%\begin{center}
% \epsfig{file=./figS4.eps,width=0.98\columnwidth}
%\end{center}
%\vskip -0.75cm
 \centerline{\includegraphics[width=0.98\columnwidth]{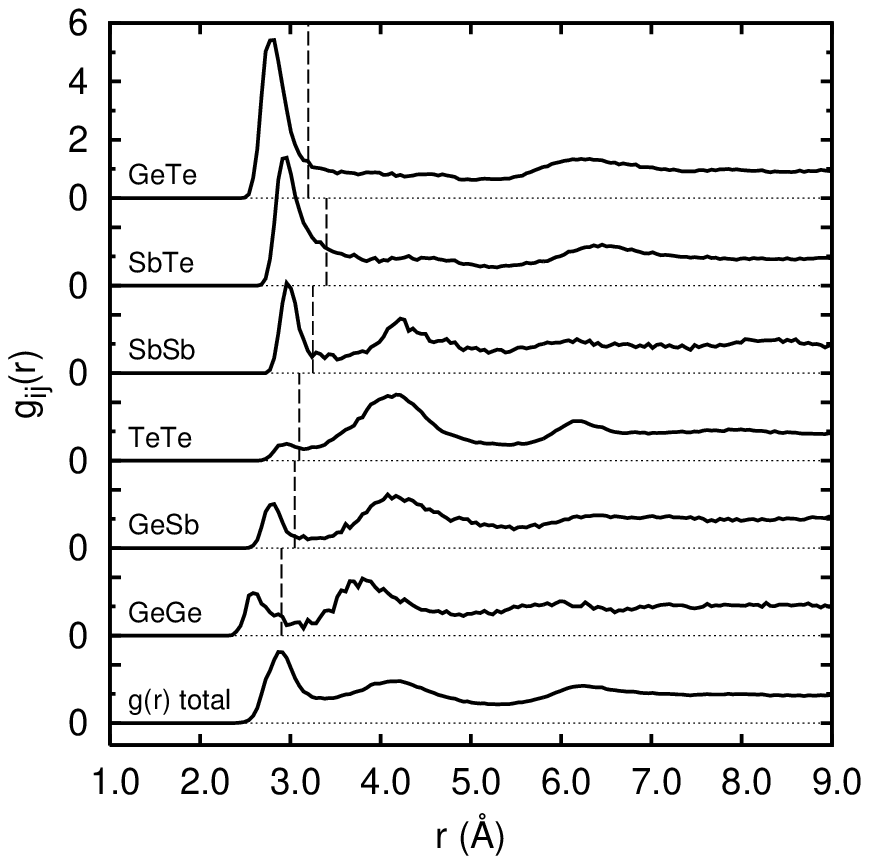}}
 \vskip -0.4cm
 \caption{Total and partial pair correlation functions of a-GST (3~ps at 300~K). The vertical lines are the
bonding cutoff used to define the coordination numbers (Table I and inset of Fig.~2). For Sb-Te the bonding
cutoff has been chosen as the outer edge of the partial pair correlation function of our model of cubic GST (300~K). The
maximum of the GeTe and SbTe pair correlation functions are 2.79 $\rm\AA$ and 2.94 $\rm\AA$ to be compared
with the experimental values of 2.61 $\rm\AA$ and 2.85 $\rm\AA$, respectively (EXAFS, Ref.~3). The first peak
in Sb-Sb and Ge-Ge partial pair correlation functions is due to homopolar bonds. We have checked that
the cutoff distances for Sb-Sb and Ge-Ge correspond indeed to the formation of  covalent bonds by looking at
the position of the  Wannier orbitals (N.~Marzari and D.~Vanderbilt, Phys. Rev. B {\bf 56}, 12847 (1997)).
The second peak in Sb-Sb and Te-Te correlation functions correspond to the diagonal of the ABAB square ring 
of the rocksalt geometry. The outer peaks above 6 $\rm\AA$ correspond to the diagonal of two planar, edge sharing 
 ABAB squares. Instead, the  peak around 5 $\rm\AA$ expected for the diagonal of the elemental rocksalt cube is 
lacking in Ge-Te and Sb-Te correlation functions. 
}
 \label{gofr}
\end{figure}
\clearpage

\renewcommand{\thefigure}{S5}
 \begin{figure}[h!]
%\begin{center}
% \epsfig{file=./figS5.eps,width=0.90\columnwidth}
%\end{center}
%\vskip -0.75cm
 \centerline{\includegraphics[width=0.98\columnwidth]{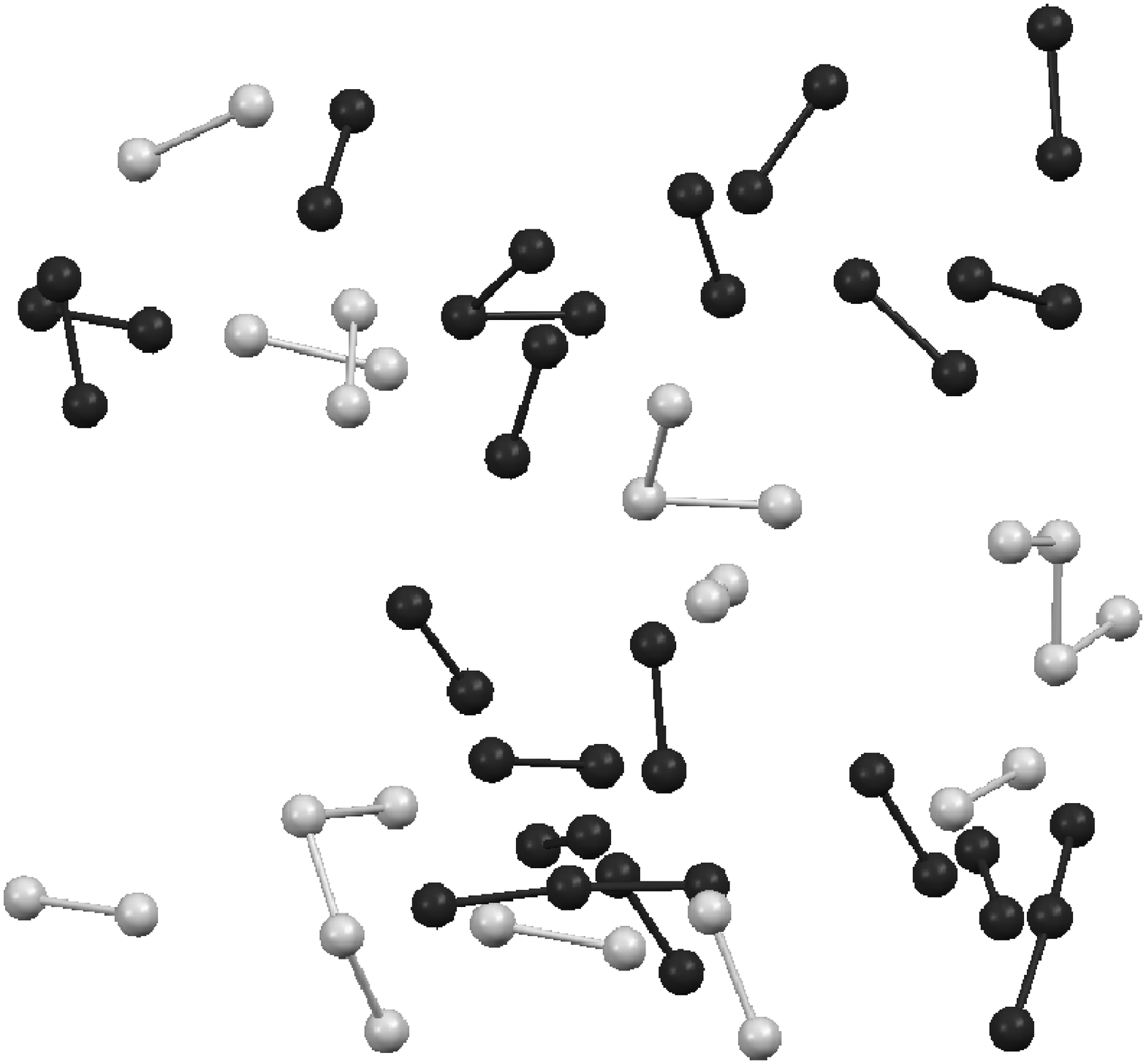}}
 \vskip -0.4cm
 \caption{Snapshot of a-GST showing Te (black spheres) and Sb (gray spheres) atoms in dimers and short chains.
}
 \label{chains}
\end{figure}

\renewcommand{\thefigure}{S6}
 \begin{figure}[ht!]
%\begin{center}
% \epsfig{file=./figS6.eps,width=0.98\columnwidth}
%\end{center}
%\vskip -0.75cm
 \centerline{\includegraphics[width=0.98\columnwidth]{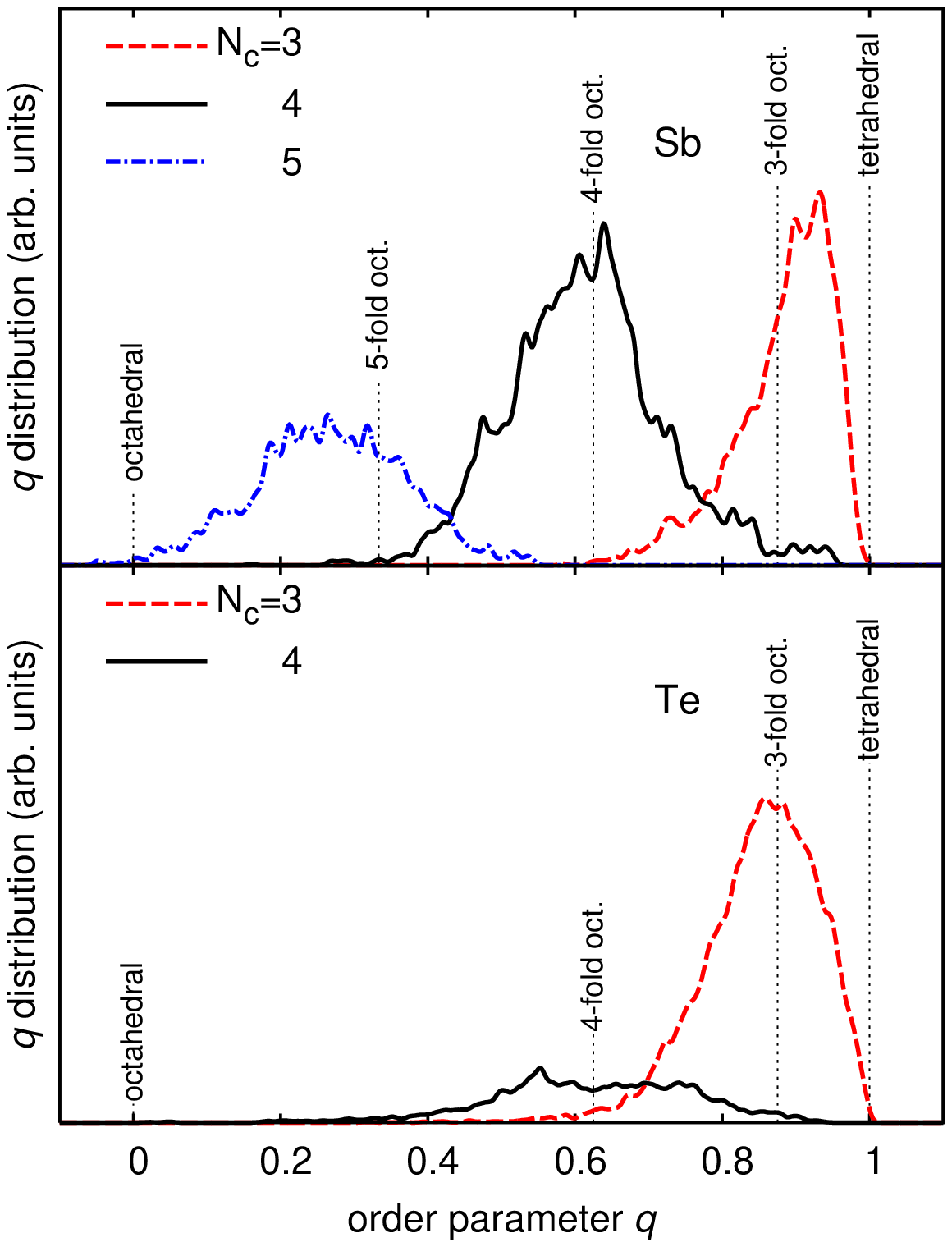}}
 \vskip -0.4cm
 \caption{Distribution of the local order parameter $q$ (see text) for Sb and Te in a-GST. Vertical lines
indicate the values of $q$ for selected ideal geometries. No Te and Sb atoms are in tetrahedral environments.
}
 \label{qSb}
\end{figure}
\renewcommand{\thefigure}{S7}
 \begin{figure}[t!]
%\begin{center}
% \epsfig{file=./figS7.eps,width=0.98\columnwidth}
%\end{center}
%\vskip -0.75cm
 \centerline{\includegraphics[width=0.98\columnwidth]{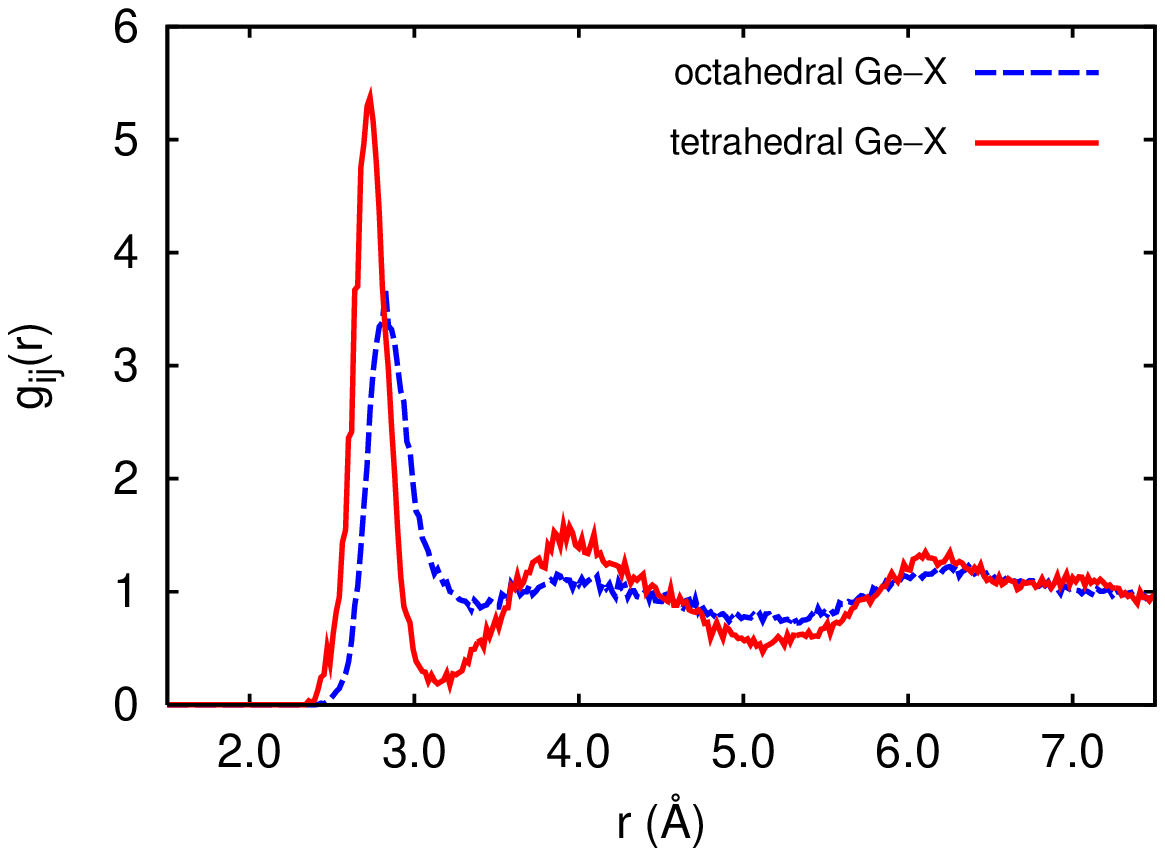}}
 \vskip -0.4cm
 \caption{Total pair correlation function for Ge in tetrahedral sites (continuous line) or in defective
octahedral sites (dashed line). Ge-X bonds are shorter for Ge in tetrahedral environment.
}
 \label{gtetra}
\end{figure}

\cleardoublepage

\end{document}